\documentstyle [12pt] {article}

\catcode`@=12
\begin{document}
\thispagestyle{empty}
\begin{flushright} DESY 98-117\\August 1998\
\end{flushright}
\mbox{}
\vspace{0.5in}
\begin{center}
{\Large	\bf Texture of a Four--Neutrino Mass Matrix\\}
\vspace{1.0in}
{\bf Subhendra Mohanty$^1$, D.P. Roy$^{2,3}$, and Utpal Sarkar$^{1,2}$\\}
\vspace{0.3in}
{$^1$ \sl Physical Research Laboratory, Ahmedabad 380 009, India\\}
\vspace{0.1in}
{$^2$ \sl DESY, Notkestrasse 85, 22607 Hamburg, Germany\\}
\vspace{0.1in}
{$^3$ \sl Tata Institute of Fundamental Research, Homi Bhabha Road, \\ }
\phantom{$^2$}{\sl Bombay - 400 005, India\\}
\vspace{1.0in}
\end{center}
\begin{abstract}\

We  propose a  simple texture  of the neutrino  mass   matrix with one
sterile neutrino along with the three standard  ones. It gives maximal
mixing  angles  for  $\nu_e  \to  \nu_S$ and  $\nu_\mu   \to \nu_\tau$
oscillations or vice versa. Thus with only  four parameters, this mass
matrix can explain  the  solar neutrino anomaly,  atmospheric neutrino
anomaly, LSND result and  the hot dark matter  of the  universe, while
satisfying all other Laboratory  constraints. Depending on the  choice
of parameters, one  can get the vacuum oscillation  or the large angle
MSW solution of the solar neutrino anomaly. 

\end{abstract} 

\newpage
\baselineskip 18pt

Recently the super-Kamiokande experiment has confirmed the atmospheric 
neutrino oscillation result, suggesting nearly maximal mixing of 
$\nu_\mu$ with another species of neutrino \cite{atm}. The same
experiment has also confirmed the solar neutrino oscillation result, 
which suggests mixing of $\nu_e$ with another species of neutrino
\cite{sol}. Moreover, the energy spectrum of the recoil electron 
seems to favour the large mixing-angle vacuum oscillation of $\nu_e$
over the MSW solutions \cite{sol}, although this may have limited 
statistical significance in the global fit to the solar neutrino data
\cite{soldata1,soldata2}. They have led to a flurry of phenomenological models
for neutrino mass and mixing which can account for these oscillations
[5-8], most of which are focussed on the bi-maximal mixing angles for the 
atmospheric and the solar neutrinos. However, almost all of these 
works are based on the three-neutrino formalism, involving the standard 
left handed neutrinos $\nu_e, \nu_\mu$ and $\nu_\tau$ \cite{nu1}.

On the other hand the inclusion of the LSND neutrino oscillation
result \cite{lsnd} is known to require a fourth neutrino, which
has to be a sterile one ($\nu_S$) for consistency with the observed
Z--width \cite{lep}. Moreover it requires either $\nu_\mu$ or $\nu_e$
to oscillate into $\nu_S$ for explaining the atmospheric and solar
neutrino anomalies, while requiring $\nu_\mu \to \nu_e$ oscillation
for the LSND result. Thus the three-neutrino models for atmospheric and 
solar neutrino anomalies, based on a $\nu_e-\nu_\mu-\nu_\tau$ mixing, 
are in direct conflict with the LSND result. While the LSND result has not
been corroborated by the preliminary KARMEN data \cite{karmen}, the
statistical significance of the latter is limited by its lower 
sensitivity in the relevant region of parameter space. Indeed, with 
the standard statistical method the 90 \% c.l. limit of KARMEN excludes 
only half the parameter space of the LSND data in the $\Delta m^2 \leq
2 eV^2$ region \cite{giunti}. Hopefully, 
this issue will be resolved by the proposed mini-BOONE experiment at 
Fermilab along with more data from KARMEN. It seems to us premature,
however, to rule out the LSND result at present. Therefore we have tried 
to construct a four-neutrino mass matrix, which can account for the present
solar and atmospheric neutrino data along with the LSND result. It can 
also account for the hot dark matter content of the universe \cite{dm},
while satisfying all laboratory and astrophysical constraints 
\cite{double,chooz,lab}.

Table 1 summarises the experimental constraints on neutrino mass and 
mixing parameters, which are relevant for our model. The large angle
MSW and the vacuum oscillation solutions to the solar neutrino data 
\cite{sol,ge,home} are taken from a recent fit to the $\nu_e$ suppression
rates along with the recoil electron spectrum by Bahcall, Krastev and
Smirnov \cite{soldata1}. For both the solutions the fit favours the 
oscillation of $\nu_e$ into a doublet neutrino
over $\nu_e \to \nu_S$. The reason is that in the  former case the 
NC scattering of this doublet neutrino with electron can partly
account for the discrepancy between the observed suppression rates 
in super-Kamiokande and the Homestake experiments. On the other hand
one can get acceptable solutions with $\nu_e \to \nu_S$ oscillation
if one makes allowance for a 20 \% 
normalisation uncertainty for the Homestake experiment. This will
also enlarge the acceptable range of $\Delta m^2$. Therefore we 
shall consider both the oscillation scenarios $\nu_e \to \nu_S$ and 
$\nu_e \to \nu_\tau$ in our model. It should be added here that the 
best value of $\sin^2 2\theta$ for the large angle MSW solution is
slightly less than 1; and even there the quality of fit is rather
poor when all the experimental data are put together \cite{soldata1}.
However one can get acceptable fit with the large angle MSW solution,
including the $\sin^2 2\theta = 1$ boundary, if one makes reasonable 
allowance for the uncertainty in the Boron neutrino flux \cite
{soldata1,soldata2}. Finally, the
global fits \cite{soldata1,soldata2} have also found acceptable small angle
MSW solutions for both these oscillation scenarios. But we do not
consider them here, since the texture of our mass matrix naturally 
leads to bimaximal mixing as we shall see below. 

\begin{table}
\caption{Present experimental constraints on neutrino masses and mixing}
\begin{center}
\begin{tabular}{||rcl||}
\hline \hline
Solar Neutrino {\cite{soldata1}}
&:& $\Delta m^2 \sim (0.8 - 2) \times 10^{-5} eV^2$ \\
(Large angle MSW) \phantom{[1]}
&& $ \sin^2 2 \theta \sim 1$ \\
Solar Neutrino {\cite{soldata1}}
&:&$ \Delta m^2 \sim (0.5 - 6) \times 10^{-10} eV^2$ \\ 
( Vacuum oscillation) \phantom{[1]} &&$  \sin^2 2 \theta \sim 1$ \\
Atmospheric Neutrino \cite{atm}
&:&$ \Delta m^2 \sim (0.5 - 6) \times 10^{-3} eV^2$ \\ 
&&$ \sin^2 2 \theta > 0.82 $\\
LSND \cite{lsnd} &:& $\Delta m_{e \mu}^2 \sim (0.4 - 2) eV^2$ \\
&& $\sin^2 2 \theta_{e \mu} \sim 10^{-3} - 10^{-2} $\\
Hot Dark Matter \cite{dm} &:& $\sum_i m_i \sim (4 - 5) eV $\\
Neutrinoless  Double  Beta  Decay \cite{double} &:&
$m_{\nu_e} < 0.46 eV$ \\
CHOOZ \cite{chooz} &:& $\Delta m_{e X}^2 < 10^{-3} eV^2 $\\
&&(or $\sin^2 2 \theta_{eX} < 0.2)$ \\
\hline \hline
\end{tabular}
\end{center}
\end{table}

The dark matter constraint on the sum of neutrino masses comes from
a recent global fit to the spectrum of density perturbation in the
universe using various cosmological models \cite{dm}. The best fit
is obtained with a hot and cold dark matter model, where the former 
constitutes 20 \%
of the critical density. Besides there is an astrophysical upper bound 
on the number of neutrino species from nucleosynthesis, which allows
1 or atmost 2 sterile neutrinos \cite{nucleo}. 

We shall consider a four-neutrino mass matrix for the three doublet 
neutrinos and a singlet (sterile) neutrino. We shall present two 
scenarios, where the solar and atmospheric neutrino anomalies are
explained by the oscillations (A) $\nu_e \to \nu_S$ and $\nu_\mu
\to \nu_\tau$ and (B) $\nu_e \to \nu_\tau$ and $\nu_\mu \to \nu_S$.
The corresponding mass matrices will be related to one another by 
suitable permutation of neutrino indices. In each case maximal 
mixing between the oscillating neutrino pairs will be ensured by the 
texture of the mass matrix. Moreover we shall obtain the vacuum 
oscillation and the large angle MSW solutions in each case depending on
the choice of parameters.

{\bf (A)  $\nu_e \to \nu_S$ and $\nu_\mu \to \nu_\tau$ Oscillations :}

In this case the texture of our neutrino mass matrix in the basis
$[\nu_e ~~ \nu_\mu ~~ \nu_\tau ~~ \nu_S]$ is
\begin{equation}
m_\nu = \pmatrix{ 0 & 0 & a & d \cr 0 & c & b & 0 \cr
a & b & 0 & 0 \cr d & 0 & 0 & 0 } .
\end{equation}
Note that it has only 4 parameters. In comparison the earlier mass
matrices considered had at least 5 parameters 
\cite{4nu1,4nu2,4nu3}. Moreover,
the above mass matrix is minimal in the sense that it has only one
diagonal element. The mass matrix of \cite{4nu1} has effectively 
4 parameters in the case of maximal vacuum oscillation solution of 
the solar neutrino. However it contains two equal diagonal elements,
which could in general be different from one another. 
It is clear from the mass matrix that the 
neutrinoless double beta decay vanishes because
$\sum_i U_{ei}^2 m_i = m_{ee} = 0$; so that the corresponding 
constraint \cite{double} is automatically satisfied.

For the $3 \times 3$ submatrix of doublet neutrinos, we shall assume the hierarchy 
\begin{equation}
b \gg a,c .
\end{equation} 
Consequently the $\nu_\mu$ and $\nu_\tau$ will form a nearly 
degenerate pair with maximal mixing and small mass squared difference
$(\sim 2 bc)$ to explain the atmospheric neutrino anomaly. Moreover,
the remaining eigenvalue of this $3 \times 3$ submatrix gets a
tiny double see-saw contribution $2 a^2 c / b^2$, which will be 
much smaller than $d$ over a wide range of the latter parameter. 
Consequently, the $\nu_e$ and $\nu_S$ will form a nearly degenerate 
pair with maximal mixing and small mass squared difference to
explain the solar neutrino anomaly. The vacuum oscillation and the 
large angle MSW solutions will correspond to the choices $d <
a,c$ and $d \sim b$ respectively. 

{\bf I -- Vacuum Oscillation Solution $(b>> a,c >d)$ :}
In this approximation the mass eigenvalues are given by,
\begin{eqnarray}
m_1 &=& d + {a^2 c \over 2 b^2} \nonumber\\
m_2 &=& b + {c \over 2} + {a^2 \over 2 b} +{c^2\over 8b}- {a^2 c \over 2 b^2} \nonumber \\
m_3 &=& -b + {c \over 2} - {a^2 \over 2 b} -{c^2\over 8b}- {a^2 c \over 2 b^2} \nonumber\\
m_4 &=& -d + {a^2 c \over 2 b^2} 
\end{eqnarray}
and the corresponding mass eigenstates ($\nu_i^T \equiv \{\nu_1 ~~ \nu_2 ~~
\nu_3 ~~ \nu_4 \}$) are related  
to the weak eigenstates ($\nu_\alpha^T \equiv \{\nu_e ~~ \nu_\mu ~~
\nu_\tau ~~ \nu_S \}$) through the mixing matrix $U_{i \alpha}$ as,
\begin{equation}
\pmatrix{\nu_e \cr \nu_\mu \cr
\nu_\tau \cr \nu_S}
 =\pmatrix{ {1 \over \sqrt{2}} & -s_1  & s_1 & 
{ 1 \over \sqrt{2} } \cr s_1 & {1 \over \sqrt{2}} &
- {1 \over \sqrt{2}} & s_1 \cr s_2^\prime & {1 \over \sqrt
{2}} &
{1 \over \sqrt{2}} & -s_2^\prime \cr - { 1 \over \sqrt{2} }
& s_2 & s_2 & {1 \over \sqrt{2}}}\pmatrix{\nu_1 \cr \nu_2 \cr
\nu_3\cr \nu_4} 
\label{U}
\end{equation}
where, $s_1 = {a \over \sqrt{2} b} $ and we can neglect the terms 
$s_2$ and $s_2^\prime$, which are of the order of $\sim O({ac \over 
\sqrt{2} b^2},{ad \over b^2}) \sim 10^{-5}$ for our choice of parameters. 
For the given $4 \times 4$ mixing matrix
$U_{i \alpha}$ the probability of two flavour oscillation is given by,
\begin{equation}
P_{\nu_\alpha \nu_\beta} = \delta_{\alpha \beta} - 4 \sum_{j > i}
U_{\alpha i} U_{\alpha j} U_{\beta i} U_{\beta j} \sin^2
\left( {\Delta m_{ij}^2 L \over 4 E} \right) ,
\label{P1}
\end{equation}
where, $\Delta m_{ij}^2 = m_i^2 - m_j^2$. For our mixing matrix the
flavour oscillation in each case is dominated by one mixing angle
which can be determined by comparing the expression 
(\ref{P1}) 
with the effective $2 \times 2$  flavour oscillation formula
\begin{equation}
P_{\nu_\alpha \nu_\beta} = 
\sin^2 2 \theta_{\alpha \beta} ~\sin^2
\left( {\Delta m_{ij}^2 L \over 4 E} \right) . 
\label{P2}
\end{equation}
Thus we get an expression for $\sin^2 2 \theta_{\alpha \beta}$ in terms 
of parameters of the mixing matrix $U_{i \alpha}$.

For illustration we shall now present the solution for a specific set of 
parameters, {\it i.e.},  $a = 0.05 eV, b = 1.5 eV, c = 0.001 eV$ 
and $d = 0.0001 eV$. 
There are two pairs of nearly degenerate  eigenvalues
\begin{eqnarray}
m_{\nu_1} \simeq - m_{\nu_4} &\simeq & d = 0.0001 eV \nonumber \\
m_{\nu_2} \simeq - m_{\nu_3} &\simeq & b = 1.5 eV. \nonumber \\
\end{eqnarray}
The LSND experiment can be explained by the oscillations between states
with mass squared difference of the order $ eV^2$ which means that it can be explained
by the oscillations between the $\nu_{1,4}$ and $\nu_{2,3}$ states. To
explain
LSND as an oscillation between the $\nu_e$ and $\nu_\mu$
 the effective mixing angle  $\sin^2 2 \theta_{e \mu}$
is obtained by comparing
(\ref{P1})
and (\ref{P2}) and reading off the mixing matrix elements from (\ref{U}),
\begin{eqnarray}
sin^2 2\theta_{e \mu}&=& - 4 \{U_{e1}U_{e2}U_{\mu1}U_{\mu 2}+
U_{e1}U_{e3}U_{\mu 1}U_{\mu 3}+U_{e4}U_{e2}U_{\mu4}U_{\mu 2}
+U_{e4}U_{e3}U_{\mu 4}U_{\mu 3}\}\nonumber\\
&=& - 4\times 4  (-s_1^2)
({1\over\sqrt{2}})^2= {4 a^2 \over b^2} .
\end{eqnarray}
Similarly the other 
 masses and the
relevant mass squared differences and the corresponding mixing 
angles for the experiments listed in Table 1 are given by,
\begin{eqnarray}
\Delta m_{sol}^2 =
 m_{\nu_1}^2 - m_{\nu_4}^2 &=& {2 a^2 c d \over b^2}
= 2.2\times 10^{-10} eV^2 \nonumber \\
\sin^2 2 \theta_{e S}&=& 1 \nonumber \\ 
\Delta m_{atm}^2 = m_{\nu_2}^2 - m_{\nu_3}^2 &=&2 b c = 0.003 eV^2
\nonumber \\ \sin^2 2 \theta_{\mu \tau}&=& 1 \nonumber \\ 
\Delta m_{LSND}^2 = m_{\nu_1}^2 - m_{\nu_2}^2 &=& b^2 -d^2 =
 2.2 eV^2 \nonumber \\ 
\sin^2 2 \theta_{e \mu}&=& 8 s_1^2 = {4 a^2  \over 
 b^2  } = 0.004 \nonumber \\
m_{DM}^{\phantom{2}} = \sum_i | m_i | &=& 3 eV .
\label{five}
\end{eqnarray}
The $\nu_e \to \nu_S$ oscillation gives the vacuum oscillation solution
to the solar neutrino anomaly, while the $\nu_\mu \to \nu_\tau$ oscillation
explains the atmospheric neutrino anomaly. The LSND result is 
explained by $\nu_\mu \to \nu_e$  oscillation. 
The contribution to dark matter is 3 eV. The CHOOZ \cite{chooz}
and other laboratory constraints are satisfied. 

Note that the above solution consists of two nearly degenerate pairs
of maximally mixed neutrinos, separated by a realatively large mass
squared gap. This is known to be the favoured mass configuration for
satisfying the various laboratory constraints \cite{4nu1,bil}.
The four model parameters are used to ensure that the three mass
squared gaps correspond to the required values of $\Delta m^2$ for
the solar, atmospheric and LSND neutrino oscillations, and $\theta_{
e \mu}$ corresponds to the required mixing angle for LSND. The hot
dark matter prediction comes out as a bonus. All these features
are natural predictions of our mass matrix; and as such they will be
shared by each of the alternative solutions discussed below. It should
also be noted that the underlying double see-saw mechanism is responsible 
for generating mass squared gaps differing by 10 orders of magnitude 
starting with mass parameters, which differ by only 3--4 orders of
magnitude ({\it i.e.}, similar to the case of the up type quark
mass matrix).

{\bf II -- Large Angle MSW Solution ($b > d >> a,c$) :} 
In this approximation ($b \neq d$)the mass eigenstates are given by,
\begin{eqnarray}
m_1 &=& d - {a^2 d \over 2(b^2-d^2)} + {a^2 b^2 c \over 2
(b^2-d^2)^2}\nonumber\\
m_2 &=& b + {c \over 2} + {c^2\over 8b}+{a^2 b \over 2(b^2 -d^2)} 
- {a^2 b^2 c \over 2 (b^2 -d^2)^2} \nonumber \\
m_3 &=& -b + {c \over 2} -{c^2\over 8b} - {a^2 b  \over 2(b^2 -d^2)} - 
{a^2 b^2 c \over 2 (b^2-d^2)^2} \nonumber\\
m_4 &=& -d + {a^2 d \over 2( b^2-d^2)}  + {a^2 b^2 c \over 2 (b^2-d^2)^2} 
\label{M2}
\end{eqnarray}
and the mixing matrix has the same form as (\ref{U}), with
 $s_1 = {a \over(b^2 + d^2)^{1/2}} $,
and $s_2=s_2^\prime = {a b \over  d (b^2 + d^2)^{1/2}} $ .
Hence like before $\nu_e \to \nu_S$ mixing and the
$\nu_\mu \to \nu_\tau$ mixing are maximal, where as the $\nu_e \to
\nu_\mu$ mixing is given by the small parameter $s_1$. It may be added
here that one gets a smooth numerical solution at $b=d$, although the
approximation (10) breaks down there.

In this case let us consider a choice, $a = 0.025 eV, 
b= 1.5 eV, c = 0.0015 eV$ and $d= 1.25 eV$. Then the different 
masses and the relevant mass squared differences and the corresponding 
mixing angles are given by,
\begin{eqnarray}
m_{\nu_1} \simeq - m_{\nu_4} &\simeq& d = 1.25 eV \nonumber \\
m_{\nu_2} \simeq - m_{\nu_3} &\simeq& b = 1.5 eV \nonumber \\
\Delta m_{sol}^2 = m_{\nu_1}^2 - m_{\nu_4}^2 &=& {2 a^2 b^2 c d \over
(b^2-d^2)^2}
= 1.1 \times 10^{-5} eV^2 \nonumber \\
\sin^2 2 \theta_{e S}&=& 1 \nonumber \\ 
\Delta m_{atm}^2 = m_{\nu_2}^2 - m_{\nu_3}^2 &=& 2 b c = 0.004 eV^2
\nonumber \\ \sin^2 2 \theta_{\mu \tau}&=& 1 \nonumber \\ 
\Delta m_{LSND}^2 = m_{\nu_1}^2 - m_{\nu_2}^2 &=& b^2 -d^2 =
 0.69 eV^2 \nonumber \\ \sin^2 2 \theta_{e \mu}&=& 8 s_1^2 =
 8 { a^2 \over (b^2 +d^2) } = 1.3 \times 10^{-3} \nonumber \\
m_{DM}^{\phantom{2}} = \sum_i |m_i| &=& 5.5 eV .
\label{seven}
\end{eqnarray}
The numerical values of the mass square differences have been calculated
using the exact solutions of for the masses. The  analytical expressions
for the mass square differences are from the polynomial approximation
(\ref{M2}).
The numerical agreement for the mass square differences between the exact
solutions and the polynomial approximation agrees upto 6 decimals in this
range of parameters.

Thus the $\nu_e \to \nu_S$ oscillation provides the large angle MSW 
solution to the solar neutrino problem.
The $\nu_\mu \to \nu_\tau$ and $\nu_\mu \to \nu_e$ oscillations
explain the atmospheric neutrino anomaly and the LSND result 
respectively as before. The contribution to dark matter is 5 eV. It
may be added here that with these parameters $s_1 < s_2$; so the effective
mass of $\nu_e$ is slightly lower than that of $\nu_S$.

{\bf (B)  $\nu_e \to \nu_\tau$ and $\nu_\mu \to \nu_S$ Oscillations :}

In this case we can use the same mass matrix as before if we make
the following change of basis,
\begin{equation}
(\nu_e ~~ \nu_\mu ~~\nu_\tau ~~ \nu_S) \longrightarrow
(\nu_e ~~ \nu_\mu ~~ \nu_S ~~\nu_\tau ) .
\end{equation}
Note that with this change of basis the diagonal element for the sterile 
neutrino $m_{SS}$ continues to remain zero, which is an important 
condition as we shall see below. Moreover the change of basis does
not affect the mass eigenvalues $m_1, m_2, m_3$ and $m_4$. But
now the nearly degenerate pair $m_1$ and $m_4$ represent the two maximally
mixed states of $\nu_e$ and $\nu_\tau$, while $m_2$ and $m_3$
represent similar admixtures of $\nu_\mu$ and $\nu_S$. Thus
the solutions (\ref{five}) and (\ref{seven}) will continue to hold with
the exchange
of the neutrino flavour indices $\tau$ and $S$ in $\theta_{e S}$
and $\theta_{\mu \tau}$. Consequently they will represent the
vacuum oscillation and large angle MSW solutions to the solar neutrino
anomaly via $\nu_e \to \nu_\tau$ oscillation, while the atmospheric
neutrino anomaly is explained via $\nu_\mu \to \nu_S$ oscillation. 
The rest of the results remain the same as before. 

Let us briefly discuss the possible mechanisms underlying the above 
mass matrix. Consider first the $3 \times 3$ submatrix corresponding 
to the three left-handed doublet neutrinos. This sub-eV scale mass
matrix could arise from the standard see-saw mechanism with three 
heavy right-handed singlet neutrinos. Alternatively, one can get it 
without any right-handed neutrino but with an expanded higgs sector via a 
radiative mechanism \cite{zee} or Majorana coupling of the left-handed
neutrino pairs to a heavy Higgs triplet \cite{ma}. In both cases one
can naturally obtain a sub-eV scale mass matrix. The extension of the 
mass matrix to include a light singlet neutrino has been tried recently 
in each of the above three models \cite{nu2,nu3,nu4}. In the standard
see-saw model it is assumed to be one of the right-handed singlets
while one adds a singlet neutrino in the other two models. In each
case one has to impose a zero Majorana mass for this singlet, as
otherwise it will naturally assume a high mass value. This is the
reason why we have set $m_{SS}$ to zero in our mass matrix (1). 
In the standard see-saw model this has been done by assuming a singular
Majorana mass matrix for the singlet neutrinos, so that one of the 
eigenvalues ($m_{SS}$) is zero \cite{nu4}. In the other two models
the $m_{SS}$ is made to vanish by imposing an additional symmetry 
\cite{nu2,nu3}. Finally one asks if this singlet neutrino can naturally 
have Dirac masses in the $\leq 1$ eV scale? It seems possible to get 
it in the Zee model \cite{nu2} and the triplet higgs model \cite{nu3}
via the same suppression mechanisms which keep the $3 \times 3$ 
doublet mass matrix in the sub-eV range. But in the case of the 
standard see-saw model it had to be put in by hand \cite{nu4}. 
We feel it is important to look for a more natural way to keep
the Dirac masses small in this model.

To summarize, we have presented a texture of four-neutrino mass matrix
which automatically ensures bi-maximal mixing between $\nu_e \to \nu_S$
and $\nu_e \to \nu_\tau$ or vice versa. Thus with only four parameters
it can account for the solar, atmospheric and LSND neutrino anomalies
while remaining consistent with other experimental constraints. The
prediction of the desired hot dark matter density comes out as 
bonus. Depending on the choice of parameters we can get both the vacuum
oscillation and the large angle MSW solutions to the solar neutrino
anomaly. Thanks to the underlying double see-saw mechanism, one can 
generate the desired mass squared gaps differing by 10 orders of magnitude
starting with the four mass parameters which differ by only 3--4 orders
of magnitude.

~\vskip 0.5in
\begin{center} {ACKNOWLEDGEMENT}
\end{center}

We are grateful to Profs.K.S.Babu and Ernest Ma and
also to Prof.K.Whisnant for pointing out a notational error in the original
version of our mass-matrix. We are also grateful to Prof. Ernest Ma for
a critical reading of the manuscript.
We thank 
Profs. V. Barger,
 W. Buchmuller, S. Goswami, S. Pakvasa, T. Weiler and P. Zerwas for
discussions. 
Two of us (US and DPR) would like to acknowledge the 
hospitality of the Theory Group, DESY and US acknowledges financial 
support from the Alexander von Humboldt Foundation. 

\newpage
\bibliographystyle{unsrt}

\end{document}